\documentstyle[12pt,twoside,fleqn,espcrc1,psfig]{article}
\title{\vspace*{-1cm}{\hfill\normalsize LBNL-40988}\\
Strangelets and Strange Quark Matter}
\author{J\"urgen Schaffner-Bielich\address{Nuclear Science Division, Lawrence
Berkeley National Laboratory,\\ University of California, Berkeley, CA 94720}
\thanks{Feodor-Lynen fellow of the Alexander-von-Humboldt Stiftung, Germany}}
\begin{document}
\maketitle

\begin{abstract}
I summarize the 
properties of finite lumps of strange quark matter (strangelets)
with emphasis on the two 
scenarios of producing strange matter in relativistic heavy ion collisions.
As an outlook, I discuss the possibility of 
short-lived strange composites and charmed matter.
\end{abstract}

\section{INTRODUCTION -- PRODUCING STRANGE MATTER IN THE LABORATORY}

As it has been discussed widely during this conference, strangeness opens a new
dimensions to nuclear physics. Insofar, systems with strangeness number
$S=-1,-2$ have been discussed. Here we want to examine the unknown domain of
finite nuclear systems with $S<-2$. There have been speculations about the
existence of finite systems of strange quark matter (strangelets) 
and strange hadronic matter.
Here we will focus on the former objects and recent progress in this field
\cite{Scha97} as the latter ones were discussed at
the last hypernuclear meeting in detail \cite{Gal92}.
How can one produce such strangeness-rich systems? 
Hadron beams enable only to explore systems
up to $S=-2$. Nevertheless, relativistic truly heavy-ion collisions constitute
a prolific source of strangeness as dozens of hyperons are produced on a single
central event. In principle, strangelets can be produced via two different
scenarios: by a coalescence of hyperons or by a distillation of a quark-gluon
           plasma. 

The coalescence model for strangelet production in heavy-ion collisions
has been put forward by Carl Dover \cite{Dover93}. The formation of a
quark-gluon plasma is not needed in this scenario. Hyperons coalesce during
the late stage of the collision forming a doorway state for strangelet
production. For example, the H dibaryon would be formed by the coalescence of
two $\Lambda$'s or a $\Xi$ with a nucleon which transform to a dibaryon with
the same quantum numbers. The production rates are proportional to two penalty
factors, one for adding a baryon number and the second one for
adding one unit of strangeness to the clusters 
\begin{equation}
P \propto q^{|A|} \cdot \lambda^{|S|}\, , \quad 
q = \frac{N_d}{N_N}\, , \quad \lambda = \frac{N_Y}{N_N} \quad .
\end{equation}
Here $N_i$ are the numbers of produced particles of the species $i$
per collision.
The penalty factors can be
extracted from experiment or estimated from phase-space arguments 
using a cascade model \cite{Raffa91,Baltz94}. 
At the AGS one finds at an energy of 14 AGeV $q\approx 0.03$, 
$\lambda\approx 0.13$, while at
the SPS at 200 AGeV it is $q\approx 0.0075$, $\lambda\approx 0.35$.
Composites with a high baryon number are more suppressed than those with a high
strangeness fraction. This feature gets more pronounced at higher bombarding
energy. 
Assuming a sensitivity of $10^{-8}$, clusters up to a baryon number of 
$A\leq 6$ can be produced via coalescence at the AGS, for SPS it is $A\leq 4$.
Hence, very low mass numbers are expected to be seen.

On the other hand, the distillation process provides the possibility for
producing strange\-lets up to $A=20-30$ \cite{Greiner91}. This assumes the
formation of a quark-gluon plasma with a high net baryon number.
Recently, it was shown that this scenarios could also hold for a quasi-baryon
free region as possibly encountered at the future heavy-ion colliders RHIC and
LHC due to large baryon number fluctuations \cite{Spieles96}.
The main idea is, that strange-antistrange quark pairs are abundantly produced
inside a quark-gluon plasma. The antistrange quark leaves the plasma and forms
a kaon with a surrounding light quarks which is ensured for a baryon-rich
regime. This enriches the plasma with a finite amount of strangeness.
Cooling by particle evaporation further enhances this effect and produces
finally a cold metastable strangelet.
Strangeness fractions of $f_S=|S|/A\geq 1$ are easily reached during the
distillation process. 

We will now discuss the properties of strangelets relevant for heavy-ion
physics, i.e.\ for low baryon numbers. 

\section{PROPERTIES OF STRANGE MATTER}

\subsection{Strangelets in its ground state}

The first speculation about the possible existence of
collapsed nuclei, was done by Bodmer \cite{Bodmer71}. He argued that another
form of matter might be more stable than ordinary nuclei. Transition to one form
to the other is suppressed by a barrier so that it does not occur during the
lifetime of the universe.
He discussed three
different forms: the nucleon model with large positive charge and a high
density ($R\approx 0.5$ fm), the general baryon model with small positive
charge and large strangeness, and the quark model with small positive charge
and hypercharge for large baryon numbers.
The second form is now called strange hadronic matter \cite{Scha93}, while the
latter one strangelets \cite{Chin79}. Note, that 
the paper lacks detailed calculation, as the MIT bag model \cite{DJJK75} as
well as the Walecka model \cite{Wal74} were only available a few years later.

Let us discuss briefly the MIT bag model. There are five different terms which
contribute to the total mass of the bag: the volume term proportional to the
bag pressure constant $B$, the purely phenomenological zero point energy,
the kinetic energy (which is a sum over single-particle energies), and the two
terms coming from the color magnetic and the color electric interaction between
the quarks. The dominant term is the volume term, so that the bag pressure of
bag parameter $B$ more or less determines the mass of a strangelet. 
The parameters can be fixed to hadron masses. Nevertheless, the value of the
bag parameter can not be fixed unambiguously. The original value of
$B^{1/4}=145$ MeV \cite{DJJK75} has to be contrasted with $B^{1/4}=235$ MeV
from a fit to charmonium levels \cite{Hasen80} which is compatible with
estimates 
from QCD sum rules \cite{SVZ79}. Furthermore, the coupling constant of the
one-gluon exchange extracted from the fit to hadron masses are so large, 
that the pressure for massless quarks gets
negative in bulk. This demonstrates that the bag model is an effective approach
with effective parameters which will change when going to larger baryon numbers
and/or strangeness. Therefore, we will study the properties of strange matter
for a variety of parameters. 

For a bag parameter of $B^{1/4}=145$ MeV, strange quark matter is absolutely
stable (this is Witten's scenario \cite{Witten84}). For bag parameters larger
than $B\geq 210$ MeV strange quark matter is unstable. In between, it is
metastable and can decay by weak interactions. 
The strange quark as a new degree of freedom lowers the Fermi energy and 
creates a global minimum located at
$f_S\approx 0.7$. This means that heavy strangelets are slightly positively
charged \cite{Farhi84}. 
The masses of strangelets can be estimated from the MIT bag model
when neglecting the color exchange contributions. One finds that finite size
effects make light strangelets with $A\leq 20$ metastable even when strange
quark matter is absolutely stable \cite{Greiner88}. 
Moreover, there exists a broad range of charges with a quite similar binding
energy. Note that the color magnetic and color electric potentials are
neglected in these calculations due to their complicated group structure. 
The lightest strangelets have been studied in the full bag model including
these two terms up to
$A=6$ \cite{Aerts78}. It turns out that all strangelet candidates for this mass
range are not bound with the well-known exception of the H dibaryon.
Hence we will focus on strangelets with masses around $A=6-30$ in the
following.

\subsection{Timescales -- strong and weak decay of strangelets}

Strangelets will not be in their ground state when being produced in a
heavy-ion collision. 
Suppose, a strangelet is created in the hot and dense matter with
some arbitrary strangeness, charge and baryon number. 
Strong interactions, the distillation process and particle
evaporation will alter the composition of a strangelet on a timescale of a few
hundred fm after the collision. The strangelet, if surviving this, will cool
down until it reaches the domain of weak interactions. Weak hadronic decay by
hadron emission takes place between $\tau = 10^{-5}-10^{-10}$ s. Strangelets
stable against strong interactions but decaying by the weak hadronic decay 
will be dubbed short-lived. Strangelets stable against weak hadronic decay will
then be subject to the weak leptonic decay happening on a timescale of
$\tau=10^{-4}-10^{-5}$ s. They are called long-lived in the following. 
The weak leptonic decay
is suppressed by phase space as it involves a three-body final state.
Note, that most
experiments are able to detect strangelets up to a lifetime of 50 ns
\cite{Ken97}. Recently, also the domain of short-lived strange matter is
accessible by new experiments as presented at this conference
\cite{Hank97,Brian97}. 

\begin{figure}[htbp]
\psfig{figure=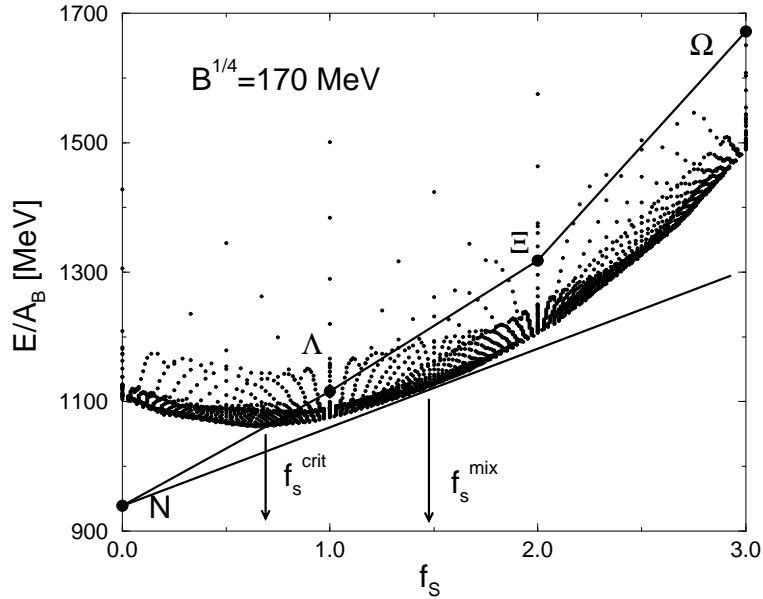,height=0.4\textheight}
\caption{Energy per baryon for strangelets up to $A=40$ using shell-mode
filling for a bag parameter of $B^{1/4}=170$ MeV. 
Note the pronounced shell effects.} 
\label{fig:strong}
\end{figure}

Figure \ref{fig:strong} shows the energy per baryon for strangelets up to
$A=40$. The line connecting the nucleon and hyperon masses stands for a free
mixture of baryons.
Strangelets with a strangeness fraction of $f_s<f_s^{crit}$ are lying
above this line. Therefore, they are unstable and decay to nucleons and
$\Lambda$'s. The tangent drawn from the nucleon mass stands for a mixture of
nucleons and strangelets. 
Strangelets in the intermediate range of
$f_s^{crit}<f_s<f_s^{mix}$ will 'move' to the tangent point increasing their
strangeness fraction to $f_s^{mix}$ by emitting nucleons as this is
energetically favourable. Hence, a strangelet will be highly charged with
strangeness when surviving the first 100 fm of the heavy-ion collision.
This is the distillation process as discussed in a dynamical approach
\cite{Greiner91}. 
Note that this chain of argumentation applies whenever strange quark matter is
metastable and when there is a minimum in
the energy per baryon at a finite $f_s$! 
Strangelets with a high strangeness fraction are most likely negatively
charged, as for isospin symmetric matter $Z/A = (1-f_s)/2$. This gives
$Z/A < -0.2$ for short-lived strangelets with $f_s\geq f_s^{mix}=1.4$.
Note also that shell effects are quite pronounced, at the order of 100 MeV/A,
which we will discuss in the next subsection in more detail.

Next, strangelets will be subject to weak hadronic decay. The two major
reactions are the weak nucleon decay ($Q\to Q' + N$) and the weak pion decay
($Q\to Q'+\pi$) by changing one unit of strangeness as discussed in
\cite{Chin79}. The change in the strangeness fraction for the weak nucleon
decay is $\Delta f_s = (f_s-1)/(A-1)$.
For a strangelet with $f_s>1$ this means
that the weak nucleon decay enhances the strangeness fraction to even higher
values. This is only true for bulk matter. Finite size effects and pockets in
the energy per baryon will certainly alter this conclusion. We want to point
out that the weak nucleon decay can only carry away positive charge. 
Only the weak neutron decay accompanied by an emission of a $\pi^-$ carries
away negative charge ($Q\to Q'+n+\pi^-$) which is suppressed by phase space.
Hence, the weak nucleon decay will also drive a strangelet to negative charge.

\subsection{Shell effects - the valley of stability}

\begin{figure}[htbp]
\psfig{figure=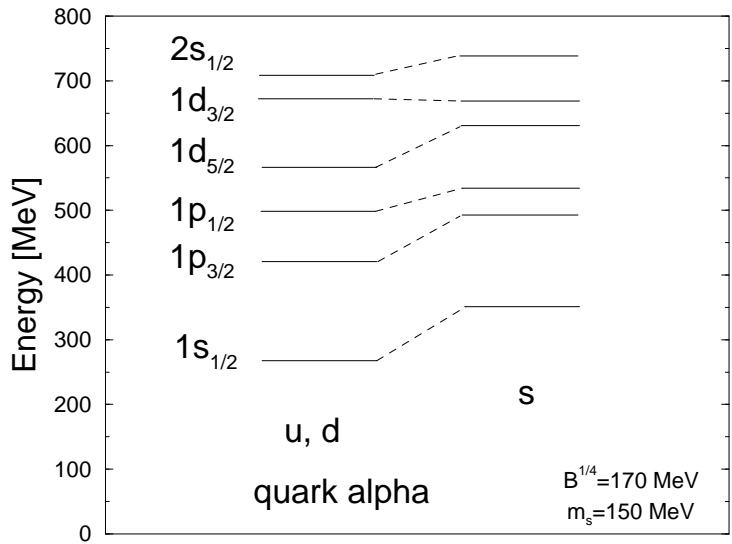,height=0.4\textheight}
\vspace*{-1cm}
\caption{The shell levels of the quark alpha for a bag parameter of $B^{1/4} =
170$ MeV.} 
\label{fig:shells}
\end{figure}

As we have seen, shell effects seems to be quite pronounced. 
Figure \ref{fig:shells} shows the single particle energy of the quark alpha
which has six up, down, and strange quarks. The lowest lying state is a 1s1/2,
which can be occupied by six quarks due to the color degree of freedom.
For the quark alpha, this state is filled up completely for all three quark
species and one can call it triple magic then. The next states are the 1p3/2
and the 1p1/2. Note that the level splitting is enormous between the 1s
and the 1p shells (more than 100 MeV!) but also the spin-orbit splitting is 
huge, about 70 MeV! Therefore, the magic numbers for strangelets 
(6, 18, and 24) will be much more pronounced than for ordinary nuclei.
The figure also shows that the single particle energy for nonstrange and
strange quarks are getting quite similar for the higher lying states as the
strange quark mass gets negligible compared to the kinetic energy.
This will allow for filling up the strange quark levels 
without loosing stability,
if the effective strange quark mass is not too high.

Magic strangelets with closed shells (triple magic) will appear irrespectively
of the bag parameter chosen. Magic strangelets with a high amount of strange
quarks and negative charge, 
as they most likely will survive a heavy ion collision, are for example
for $N_u=N_d=6$, $N_s=18,24$ and $N_u=6$, $N_d=18$, $N_s=18,24$.
It is interesting to look at the masses and charges of these four strangelets:
$A=10$ with $Z=-4$, 
$A=12$ with $Z=-6$,
$A=14$ with $Z=-8$, and
$A=16$ with $Z=-10$.
These candidates form a valley of stability for strangelets and
are highly negatively charged. 

\begin{figure}[htbp]
\psfig{figure=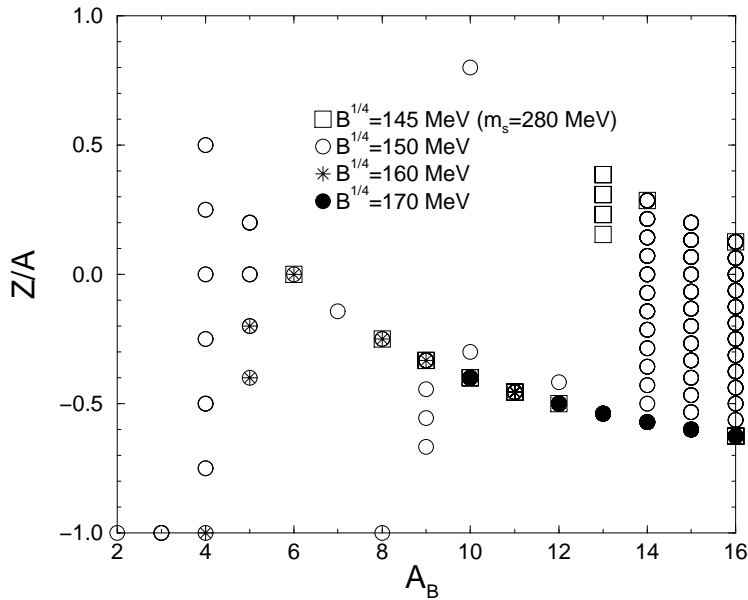,height=0.4\textheight}
\vspace*{-1cm}
\caption{Long-lived strangelet candidates and their masses and charge-to-mass
rations. The valley of stability is clearly visible.}
\label{fig:longlived}
\end{figure}

A detailed calculation in the MIT bag model including shell effects and all the
hadronic decay channels has been done in \cite{Scha97}.
It was shown that shell effects are important for long-lived candidates which
are stable against weak hadronic decay. The above listed strangelet candidates
turn out to be long-lived for most of the parametrizations used. 
Figure \ref{fig:longlived} shows the result of this investigation. 
Different bag parameters have been chosen but the valley of stability remains.
For bag parameters of $B^{1/4}=180$ MeV or larger, no long-lived candidates
have been found at all. In these case, strange quark matter starts getting
unstable against hadron emission already in bulk. 
The candidates will live
on a timescale of the weak leptonic decay $\tau=10^{-4}-10^{-5}$ s and have a
charge-to-mass ratio close to an antideuteron $Z/A\approx -1/2$.

We conclude, that 
\begin{itemize}
\item if the relativistic shell model is used, and
\item if strange quark matter is at least metastable
\item and has a local minimum at some strangeness fraction
\end{itemize}
then a valley of stability appears at $A=10-16$ with $Z\leq -4$.
If that is correct, then
it has important impacts for the present searches for strangelets in
heavy-ion collisions. Presently, the experiments focus on objects with a small
positive charge 
which is only true for heavy strangelets. Nevertheless, experiments like E864 
are also able to detect these highly charged strangelet candidates with their
present setup \cite{Ken97}.

\section{OUTLOOK}

\subsection{Charm Matter}

Insofar, we discussed the properties of quark matter including the strange
quark. What happens when going to the next heavier quark, the charm quark with
a mass of $m_c\approx 1.5$ GeV? Compared to the lighter quarks, the kinetic
energy of the charm quark will be dominated by its mass. Moreover, the charm
quark is heavier than the Fermi levels, therefore absolute stable charm matter
most likely does not exist and will decay by weak interactions on the timescale
of the lifetime of charmed hadrons, $\tau\approx 10^{-12}$ s. 
Nevertheless, there have been speculations about the existence of the
Pentaquark with 4 light quarks (u-,d-, or s-quarks) and one anticharm quark
\cite{Lipkin87}. Due to color magnetic interactions between the light quarks,
the Pentaquark might be bound. 
Most recently, charmlets have been studied in a
modified bag model \cite{SV97}. The charm quark feels a strong attractive
potential from the color electric term, while the light quarks are bound by
color magnetic forces. This two effects enable charm matter with strange quarks
to be bound even for the case of $B^{1/4}=235$ MeV where pure strange quark
matter is unstable. 

Estimates for the production of charmed matter in heavy-ion collisions have
been done by Carl Dover \cite{Dover90}. He estimates in a coalescence model 
that every $10^5$ event at
SPS might be able to produce a Pentaquark state. He also demonstrates, that
charm quarks are abundantly produced at the heavy-ion collider RHIC.
Coalescence estimates in the same spirit done in \cite{SV97} shows that 
charmlets with $A+|C|\leq 4$ might be produced at every $10^6$ event at 
RHIC. This opens the exciting perspective of probing the nuclear chart into a
fourth dimensions besides baryon number, isospin and strangeness!

\subsection{Dihyperons}

Recently, it became possible to detect in heavy-ion collisions at the AGS 
(the EOS experiments E895, E910 \cite{Brian97}) as
well as at SPS (NA49)
directly weakly decaying short-lived
hadrons by their decay topology in time projection chambers.
In addition, one experiment (E896) is especially designed to search for
short-lived strange matter \cite{Hank97}.
They are able to detect short-lived $S< -2$ systems in the laboratory for
the first time. Let us discuss in the following strange composites with $A=2$. 

A possible bound state of ($\Xi^0$p)$_b$ can decay by weak nonmesonic decay to
a $\Lambda$ and a proton. The decay topology is exactly the same for the decay
of a $\Xi^-$ which has been seen in the TPC's already 
but with the opposite charge. A bound state of $(\Xi^0\Lambda)_b$ might decay
to two $\Lambda$'s. A study of $\Lambda\Lambda$ correlations, which is
under investigation \cite{Brian97},
or backtracking techniques will reveal this exotic state. 
There exists a lot of other candidates which will have decay patters distinct
from ordinary hadrons. As E896 is also able to see neutrons \cite{Hank97},
a decay pattern of $(\Lambda\Xi^-)_b\to \Sigma^- + \Lambda$ will be measurable
in future runs. 

The production rates for dihyperons are quite high. Coalescence estimates using
the values given in this paper gives 0.2 and 0.03 dihyperons per central
AuAu collision at the AGS for $S=-2$ and $S=-3$, respectively. At the SPS one
gets 0.4 and 0.1 per event for $S=-2$ and $S=-3$, respectively. 

Even if dihyperons are not bound, correlation measurements of hyperons will
reveal important information about the hyperon-hyperon interactions. Note that
the three double $\Lambda$ hypernuclear events constitutes the only information
about the hyperon-hyperon interaction so far! Any information about this would
be of uttermost importance for our knowledge about the interaction
of baryons and the underlying symmetry (for example, does it fulfill
approximately SU(3) symmetry ?). Carl Dover studied the
hyperon-hyperon interaction using the Nijmegen model D \cite{Scha94}.
He found that the potentials between the hyperons is strongly attractive. 
Nevertheless, the existence or nonexistence 
of bound dihyperon states depends crucial on the value of the cutoff radius
which is unknown. Experiments looking into this hitherto unexplored
domain of nuclear physics are therefore impatiently awaited.

\section*{Acknowledgments}

This work is dedicated to late Carl B. Dover. 
As the outlook into the domain of multiply strange nuclear systems and into the
domain of charm matter has demonstrated, Carl Dover has done it already and
paved the way to new physics for future generations. 
I owe him a most fruitful collaboration and many
helpful and illuminating conversations from which I learned 
a lot.
J.S.B. acknowledges support
by the Alexander-von-Humboldt Stiftung with a Feodor-Lynen fellowship.
This work is supported in part by 
the Director, Office of Energy Research,
Office of High Energy and Nuclear Physics, Nuclear Physics Division of the
U.S. Department of Energy under Contract No.\ DE-AC03-76SF00098.


\end{document}